\documentclass[10pt, conference, letterpaper]{IEEEtran}

\ifCLASSINFOpdf
   \usepackage[pdftex]{graphicx}
   \graphicspath{{../pdf/}{../jpeg/}}
   \DeclareGraphicsExtensions{.pdf,.jpeg,.png}
\else
   \usepackage[dvips]{graphicx}
   \graphicspath{{../eps/}}
   \DeclareGraphicsExtensions{.eps}
\fi

\usepackage{amsmath}
\usepackage{amsfonts}
\usepackage{marvosym}
\usepackage{makeidx}
\usepackage{mathrsfs}
\usepackage{url}
\usepackage{graphicx}
\usepackage{algorithm}
\usepackage{algorithmic}
\usepackage[english]{babel}
\usepackage{fullpage}
\usepackage{setspace}

\usepackage{graphicx}
\DeclareGraphicsExtensions{.jpg}
\usepackage{color}
\newcommand{\be}{\begin{equation}}
\newcommand{\ee}{\end{equation}}

\begin{document}

\title{Lightweight compression with encryption based on Asymmetric Numeral Systems}

\author{\IEEEauthorblockN{Jarek Duda}
\IEEEauthorblockA{Jagiellonian University\\
Golebia 24, 31-007 Krakow, Poland\\
Email: \emph{dudajar@gmail.com}}
\and
\IEEEauthorblockN{Marcin Niemiec}
\IEEEauthorblockA{AGH University of Science and Technology\\
Mickiewicza 30, 30-059 Krakow, Poland\\
Email: \emph{niemiec@kt.agh.edu.pl}}
}

\maketitle

\begin{abstract}
Data compression combined with effective encryption is a common requirement of data storage and transmission. Low cost of these operations is often a high priority in order to increase transmission speed and reduce power usage. This requirement is crucial for battery-powered devices with limited resources, such as autonomous remote sensors or implants. Well-known and popular encryption techniques are frequently too expensive. This problem is on the increase as machine-to-machine communication and the Internet of Things are becoming a reality. Therefore, there is growing demand for finding trade-offs between security, cost and performance in lightweight cryptography. This article discusses Asymmetric Numeral Systems -- an innovative approach to entropy coding which can be used for compression with encryption. It provides compression ratio comparable with arithmetic coding at similar speed as Huffman coding, hence, this coding is starting to replace them in new compressors. Additionally, by perturbing its coding tables, the Asymmetric Numeral System makes it possible to simultaneously encrypt the encoded message at nearly no additional cost. The article introduces this approach and analyzes its security level.
The basic application is reducing the number of rounds of some cipher used on ANS-compressed data, or completely removing additional encryption layer if reaching a satisfactory protection level.
\end{abstract}



\IEEEpeerreviewmaketitle

\section{Introduction}

Reliable and efficient data transmission is a crucial aim of communications. Modern telecommunication systems are facing a new challenge: security. Usually, data confidentiality is implemented by additional services, which are able to protect sensitive data against disclosure. Unfortunately, cryptographic algorithms decrease performance. Moreover, it is impossible to implement security services in many systems with limited resources (e.g., the Internet of Things). Therefore, system architects must find other ways to ensure data protection. One such possibility is integration of encryption with other data processing steps, such as source coding.

Prefix codes, such as the well-known Huffman coding~\cite{HC}, Golomb, Elias, unary and many others, are the basis of data storage and transmission due to their low cost. They directly translate a symbol into a bit sequence. As the symbol of probability $p$ generally contains $\lg(1/p)$ bits of information $(\lg\equiv \log_2)$, prefix codes are perfect for probabilities with a power of $1/2$. However, this assumption is rarely true in practice. While encoding a sequence of $\{p_s\}$ probability distribution with a coding optimal for $\{q_s\}$ distribution, we use asymptotically $\Delta H=\sum_s p_s \lg(p_s/q_s)$ more bits/symbol than required. This cost of inaccuracy is especially significant for highly probable symbols. They can carry nearly 0 bit/symbol of information, while prefix codes have to use at least 1 bit/symbol.

Arithmetic and range coding (\cite{ari,ran}) avoid this cost by operating on nearly accurate probabilities. However, they are more costly and usually require multiplication, which is an operation with a high computational complexity. Using lookup tables to avoid multiplication is achieved for example by CABAC~\cite{CABAC} in H.264, H.265 video compressors. However, it operates on the binary alphabet, requiring eight steps to process a byte.

\begin{figure}[t!]
    \centering
        \includegraphics[width=8cm]{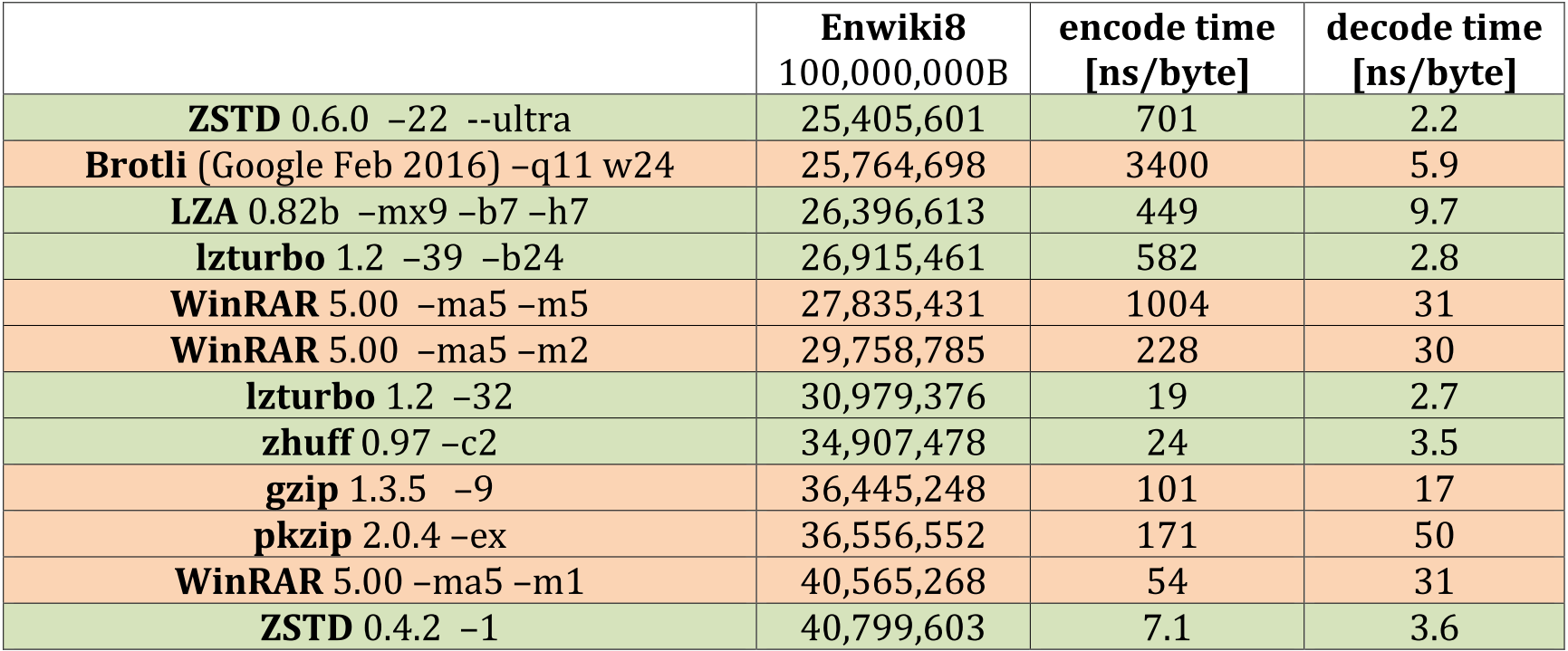}\
\begin{center}
        \caption{Comparison of some well known compressors based on Huffman coding (marked red) with those using ANS (marked green) from \cite{mah} benchmark. ZSTD, lzturbo and zhuff use the tANS variant, which allows to add encryption by perturbing coding tables. }
   \label{tab}
\end{center}
\end{figure}
Recently, a new multiplication-free large alphabet entropy coder was proposed for low cost systems: Asymmetric Numeral Systems (ANS) (\cite{ans,last,pcs2015}). In contrast to prefix codes, this coding uses nearly accurate probabilities for coded symbols. The high performance and efficiency of ANS is leading to Huffman and Range being replaced in new compressors (\cite{zhuff,lzturbo,LZA}), including Apple LZFSE~\cite{LZFSE} and Facebook Zstandard~\cite{ZSTD} to improve performance. Figure \ref{tab} presents a comparison of well known compressors based on Huffman coding with compressors using the ANS algorithm. It shows that this new entropy coder allows for compressors which are many times faster both decoding and encoding for comparable compression ratios. One of the reasons is that, while Huffman coding requires costly sorting of symbols to build the prefix tree, ANS initialization is cheap: with linear time and memory cost. This advantage is especially important for the cost of hardware implementations, which improvements have been already demonstrated for FPGA~\cite{FPGA}.

As well as providing effective data compression, another basic requirement of data storage and transmission is confidentiality. We are able to ensure data confidentiality using symmetric ciphers (asymmetric cryptography is not an appropriate solution in environments with limited resources because of its high computational cost). However, popular symmetric ciphers such as the Advanced Encryption Standard (AES), turn out to be too costly for many applications, especially battery-powered, such as autonomous remote sensors or the Internet of Things. As such, there is a growing field of lightweight cryptography~\cite{light1,light2,light3} -- with a focus on low cost, at a trade-off for having lower protection requirements.

Since a combination of compression and encryption is a common requirement, the cost priority suggests a natural solution of combining these two steps. Many approaches were considered for adding encryption into methods which are already a part of data compressors: Lempel-Ziv substitution schemes~\cite{eLZ1,eLZ2}, Burrows-Wheeler transform~\cite{eBWT} and arithmetic coding~\cite{earith1,earith2}. These articles contain some general techniques, which addition might be considered to improve security of discussed ANS-based encryption.

Huffman coding has also been discussed for adding simultaneous encryption~\cite{HCcrypt}. An abstract of an article by Ronald Rivest et al.~\cite{rivest} concludes that: \emph{"We find that a Huffman code can be surprisingly difficult to cryptanalyze"}. The main problem is the lack of synchronization -- the attacker does not know how to split the bit sequence into blocks corresponding to symbols. Additionally, data compression offers auxiliary protection by reducing redundancy which could be used for cryptanalysis.

A Huffman decoder can be viewed as a special case of the tabled variant of an ANS decoder, referred as tANS \cite{last}. This generalization allows for more complex behavior and other features, which suggest that secure encryption could be included inside the entropy coding process. While the prefix code is a set of rules: "symbol $\to$ bit sequence", tANS also has a hidden internal state $x\in \{2^R, ... ,2^{R+1}-1\}$ for some $R\in\mathbb{N}$, which acts as a buffer containing $\lg(x)\in [R,R+1)$ bits of information. The transition rules have the form $$(\textrm{symbol, state})\quad\to\quad (\textrm{bit sequence, new state})$$
Therefore, in comparison  with Huffman coding, there is an additional hidden variable $x$, which controls the bit sequence to produce, including the number of produced bits in this step: floor or ceiling of $\lg(1/p)$. As chaos is seen as strongly linked to the security of cryptography~\cite{chaos1,chaos2}, the authors discuss three sources of chaos in evolution of this internal state, making its behavior virtually unpredictable while incomplete knowledge.

As only a few ciphers like one-time pad can be formally proven to be safe, practical encryption schemes often require time to gain trust as being secure: by lack of successful attacks. Hence, while there are some arguments of strength of the proposed encryption scheme, until gaining such trust it is suggested to be used together with a convenient cipher like AES, for example with a reduced number of rounds. Comparing Huffman-based compression plus 10 round of AES, with tANS-based compression+encryption plus 5 rounds of AES, we get gain in both compression ratio and performance.

The remainder of the paper proceeds as follows. Section II introduces the ANS algorithm: coding and decoding as well as some examples of these steps. Section III, presents the basic concept of including encryption in tANS, and properties influencing security level: set of cryptographic keys, chaotic behavior, etc. Section IV describes the security features of this encryption method. Finally, Section V concludes the paper.

\begin{figure*}
    \centering
        \includegraphics[width=15cm]{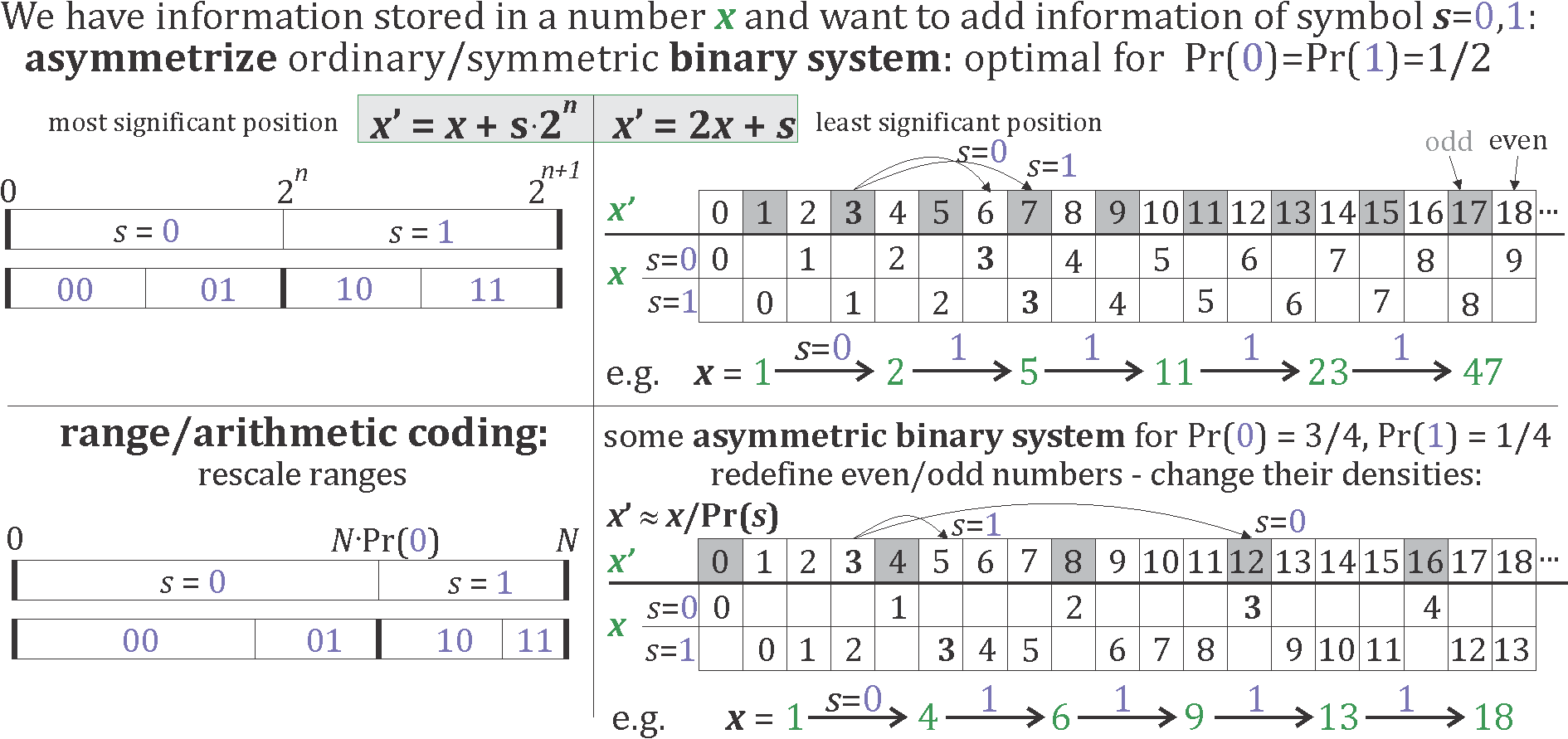}\
\begin{center}
        \caption{Arithmetic coding (left) and ANS (right) seen as an asymmetrization of the standard numeral system - in order to optimize them for storing symbols from a general probability distribution.}
        \label{ans1}
\end{center}
\end{figure*}

\section{Asymmetric Numeral Systems (ANS)}

This section introduce ANS, focusing on the tabled variant (tANS). Further discussion and other variants of ANS can be found in~\cite{last}.

\subsection{Coding into a large natural number}
Let us first consider the standard binary numeral system. It allows to encode a finite sequence of symbols from the binary alphabet ($s_i\in\mathcal{A}=\{0,1\}$) into $x=\sum_{i=0}^{n-i} s_i 2^i\in\mathbb{N}$. This number can be finally written as length $\approx \lg(x)$ bit sequence. This length does not depend on exact values of symbols -- this approach is optimized for $\Pr(0)=\Pr(1)=1/2$ symmetric case, when both symbols carry 1 bit of information. In contrast, a $\{p_s\}$ probability distribution symbol sequence carries asymptotically $\sum_s p_s \lg(1/p_s)$ bits/symbol (Shannon entropy), and a general symbol of probability $p$ carries $\lg(1/p)$ bits of information. Hence, to add information stored in a natural number $x$ to information from a symbol of probability $p$, the total amount of information will be
$\approx \lg(x)+\lg(1/p)=\lg(x/p)$ bits of information. This means that the new number $x'\in\mathbb{N}$ containing both information should be approximately $x'\approx x/p$, which is the basic concept of ANS.

Having a symbol sequence encoded as $x=\sum_{i=0}^{n-i} s_i 2^i$, we can add information from a new symbol $s\in\mathcal{A}$ in two positions: the most or the least significant. The former means that the new number containing both information is $x'=x+s 2^n$. The symbol $s$ chooses between two large ranges for $x'$: $\{0,...,2^n-1\}$ and $\{2^n, ..., 2^{n+1}-1\}$. The symmetry of their lengths corresponds to the symmetry of informational content of both symbols. As depicted in the left panel of Figure \ref{ans1}, arithmetic or range coding can be viewed as an asymmetrization of this approach to make it optimal for different probability distributions. They require operating on two numbers, defining the currently considered range, what is analogues to the need to remember the current position $n$ in the standard numeral system.

We can avoid this inconvenience by adding a new symbol in the least significant position: $C(s,x)=x'=2x+s$. Old digits are shifted one position up. To reverse this process, the decoding function is $D(x')=(s,x)=(\mod(x',2), \lfloor x'/2\rfloor)$.
This approach can be viewed that $x'$ is $x$-th appearance of an even $(s=0)$ or odd $(s=1)$ number. We can use this rule to asymmetrize this approach to make it optimal for a different probability distribution. In order to achieve this, we need to redefine the division of natural numbers into even and odd numbers, such that they are still distributed uniformly, but with the density corresponding to the assumed probability distribution. More formally, for a probability distribution $\{p_s\}$ we need to define a \emph{symbol distribution} $\overline{s}:\mathbb{N}\to \mathcal{A}$, such that: $|\{0\leq x<x':\overline{s}(x)=s\}| \approx x' p_s$. Then the encoding rule is
$$x'=C(s,x) \textrm{ is $x$-th appearance of symbol }s.$$
and correspondingly for the decoding function $D$, such that $D(C(s,x))=(s,x)$. The decoded symbol is $\overline{s}(x')$ and $x$ is the number of appearance of this symbol. More formally:
$$ C(s,x) = x' : \overline{s}(x')=s,\ |\{0\leq y<x':\overline{s}(y)=s\}|=x $$
$$ D(x')=(\overline{s}(x'),\ |\{0\leq y<x':\ \overline{s}(y)=\overline{s}(x') \}|) $$

The right panel of Figure \ref{ans1} depicts an example of such a process for the $\Pr(0)=3/4,\ \Pr(1)=1/4$ probability distribution. Starting with $x=1$ symbol/state, we encode successive symbols: 01111 into $x=47$ or $x=18$. Then we can can successively use the decoding function $D$ to decode the symbol sequence in the reverse order. ANS results in a lower representation than the standard numeral system, since it better corresponds with the digit distribution of the input sequence 01111.

There can be found arithmetic formulas using multiplication for such coding/decoding functions: uABS and rABS variants for the binary alphabet, and rANS variant for any large alphabet~\cite{last}. The range variant (rANS) can be viewed as a direct alternative to range coding with some better performance properties such as a single multiplication per symbol instead of two, leading to many times faster implementations~\cite{ryg}. However, since it requires multiplication and is not suited for encryption, this paper only discusses the tabled variant (tANS), in which we put the entire coding or decoding function for a range $x\in I$ into a table.
\begin{figure*}
    \centering
        \includegraphics[width=15cm]{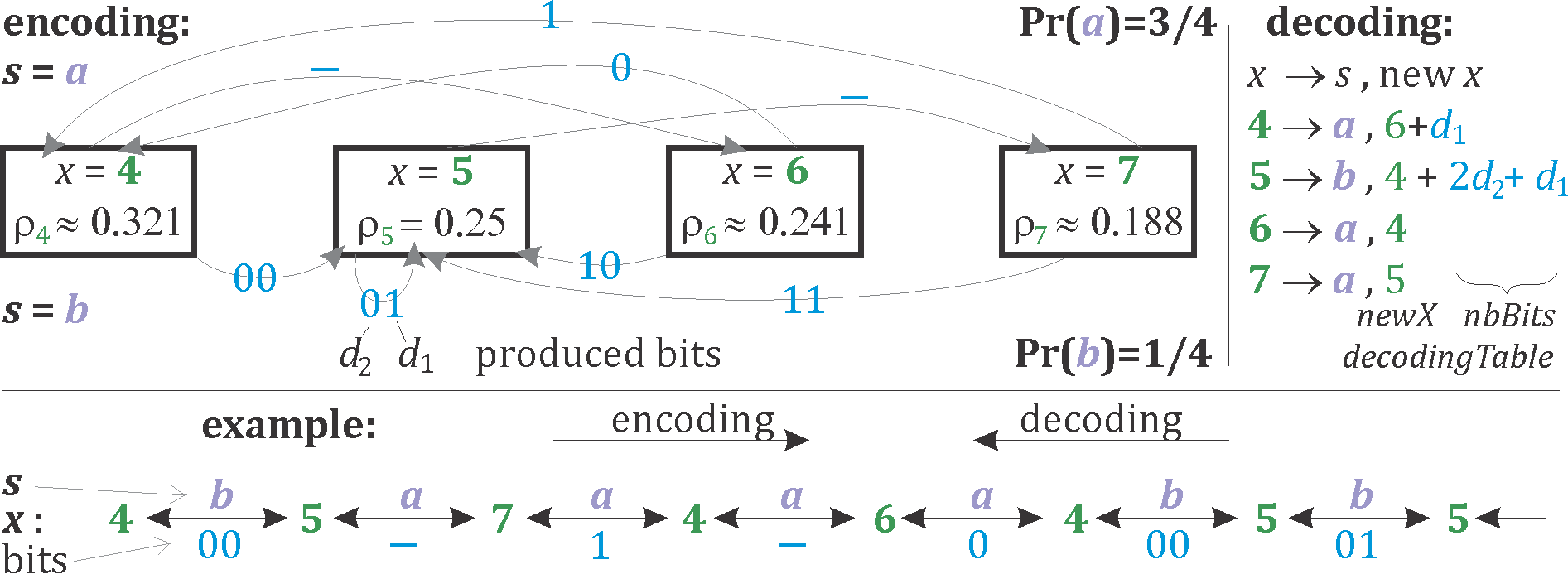}\
\begin{center}
        \caption{Example of 4 state tANS (top) and its application for stream coding (bottom). State/buffer $x$ contains $\lg(x)\in [2,3)$ bits of information.
        Symbol $b$ carries 2 bits of information, while $a$ carries less than 1 - its information is gathered in $x$ until it accumulates to a complete bit of information. $\rho_x$ are probabilities of visiting state $x$ assuming the i.i.d. input source.}
   \label{automm}
\end{center}
\end{figure*}

\subsection{Streaming ANS via Renormalization} \label{strsec}

Using the $C$ function multiple times allows us to encode a symbol sequence into a large number $x$. Working with such a large number would be highly demanding. In AC, renormalization is used to allow finite precision, an analogous approach should be used for ANS. Specifically, we enforce $x$ to remain in a fixed range $I$ by transferring the least significant bits to the stream (we could transfer a few at once, but this is not convenient for tANS). A basic scheme for the decoding/encoding step with included renormalization is:

\begin{algorithm}[htbp]
\footnotesize{
\caption{ANS decoding step from state $x$}
\label{dec1}
\begin{algorithmic}
\STATE $(s,x)=D(x)$     \qquad\qquad\qquad\quad   \COMMENT{ the proper decoding function }
\STATE useSymbol($s$)    \qquad\qquad\qquad\qquad   \COMMENT{ use or store decoded symbol }
\WHILE {$x<L$}
\STATE $x = 2\cdot x +$ readBit() \qquad\qquad \COMMENT{read bits until returning to $I$}
\ENDWHILE
\end{algorithmic}
}
\end{algorithm}

\begin{algorithm}[htbp]
\footnotesize{
\caption{ANS encoding of symbol $s$ from state $x$}
\label{enc1}
\begin{algorithmic}
\STATE \textbf{while} $x > maxX[s]$ \textbf{do} \qquad\quad \COMMENT{$maxX[s]$ will be found later}
\STATE \quad\ {writeBit$(\textrm{mod}(x,2));\ x = \lfloor x/2 \rfloor$} \qquad\COMMENT{write youngest bits}
\STATE \textbf{end while} \qquad\qquad\qquad\qquad\qquad \COMMENT{until we can encode symbol}
\STATE $x=C(s,x)$ \qquad\qquad\qquad\qquad   \COMMENT{ the proper encoding function }
\end{algorithmic}
}
\end{algorithm}

To ensure that these steps are the inverse of each other, we need to make sure that the loops for writing and reading digits end up with the same values. For this purpose, let us observe that if a range has the form $I=\{L,\ldots,2L-1\}$, when removing ($x\to \lfloor x/2 \rfloor$) or adding ($x\to 2x + d$) the least significant bits, there is exactly one way of achieving range $I$.
For uniqueness of the loop in Method~\ref{dec1}, we need to use $I$ range of this type: $I=\{L,\ldots,2L-1\}$ where for practical reasons we will use $L=2^R$.  For uniqueness of the loop in Method~\ref{enc1} we need to additionally assume that
$$I_s=\{x:C(s,x)\in I\}\qquad \left( I=\bigcup_s C(s,I_s)\right)$$
are also of this form: $I_s=\{L_s,\ldots,2L_s-1\}$ and therefore $maxX[s]=2L_s-1$ which is used in Method~\ref{enc1}.
\subsection{Tabled variant (tANS)}
In the tabled variant (tANS), which is used in most of compressors in Fig. \ref{tab} and is interesting for cryptographic purposes, we put the entire behavior into a lookup table. Let us start with the following example: we construct an $L=4$ state automaton optimized for the $\Pr(a)=3/4,\ \Pr(b)=1/4$ binary alphabet, depicted in Figure \ref{automm}. We need to choose a symbol distribution $\overline{s}:I\to\{a,b\}$ for $I=\{4,5,6,7\}$. To correspond to the probability distribution, the number of symbol appearances should be chosen as $L_a=3$, $L_b=1$. There now remain four options to choose the $\overline{s}$ function. Let us focus on the choice $\overline{s}(5)=b$, $\overline{s}(4)=\overline{s}(6)=\overline{s}(7)=a$, or in other words: "abaa" symbol spread. We need to enumerate the appearances using the numbers $I_a=\{3,4,5\},\ I_b=\{1\}$, getting the decoding function $D(4)=(a,3),\ D(5)=(b,1),\ D(6)=(a,4),\ D(7)=(a,5)$. It allows us to obtain the decoded symbol and a new state. However, some of these states are below the $I$ range, therefore we need to apply renormalization by reading some youngest bits to return to $I$ range. For example for $x=5$, the decoding function takes us to $x=1$, so we need to read two bits from the stream ($d_1,\ d_2$) to return to $I$, leading to state $x=4+2d_2+d_1$.

Assuming the input source is i.i.d. $\Pr(a)=3/4,\ \Pr(b)=1/4$, we can find the probability distribution of the visiting states of such an automaton: $\rho_x$ in this figure. It allows us to find the expected number of bits/symbol: $H'\approx 1\cdot 1/4 \cdot 2 + (0.241+0.188)\cdot 3/4\cdot 1\approx H+0.01$ bits/symbol, where $H=\sum_s p_s \lg(1/p_s)$ is the minimal value (Shannon entropy). Generally, as discussed in~\cite{last}, $\Delta H=H'-H$ behave approximately like $m^2/L^2$, where $m$ is the size of the alphabet.
\paragraph{Connection with prefix codes}
Using lookup tables, the decoding procedure can be written as:
\begin{algorithm}[htbp]
\footnotesize{
\caption{Decoding step for prefix codes and tANS}
\label{dec0}
\begin{algorithmic}
\STATE $t = decodingTable[X]$  \qquad     \COMMENT{$X\in \{0,..,2^R-1\}$ is current state}
\STATE useSymbol($t.symbol$)    \qquad\qquad   \COMMENT{ use or store decoded symbol }
\STATE $X = t.newX + $readBits$(t.nbBits)$   \qquad\qquad  \COMMENT{ state transition }
\end{algorithmic}
}
\end{algorithm}

where $X=x-L\in \{0,..,2^R-1\}$ is a more convenient representation. It should be noted that this method can also be used for decoding prefix codes such as Huffman coding. In this case $R$ should be chosen as the maximal length of the bit sequence corresponding to a symbol. The state $X$ should be viewed as a buffer containing the last $R$ bits to process. It directly determines the symbol, which uses $nbBits\leq R$ bits of the buffer. The remaining bits should be shifted and $nbBits$ should be read from the stream to refill the buffer:
$$decodingTable[X].newX = (X << nbBits)\ \&\ mask$$
where $<<$ denotes left bit-shift operation and $\&mask$ denotes restriction to the least significant $R$ bits.

Just shifting the unused bits corresponds to assuming that the produced symbol carried indeed $nbBits$ bits of information: has $2^{-nbBits}$ probability. tANS works on fractional amounts of bits by not only shifting the unused bits, but also modifying them according to the fractional amount of bits of information.

It should be noted that if we choose $L_s=2^{r_s}$ for symbol of probability $\approx 2^{r_s-R}$, and spread symbols in ranges, our tANS decoder would become a decoder for a prefix code. For example, "aaaabcdd" symbol spread would lead to decoder for $a\to 0,\ b\to 100,\ c\to 101,\ d\to 11$ prefix code. Therefore, prefix codes can be regarded as a degenerated case of tANS.

\subsection{Algorithms (\lowercase{t}ANS)} \label{tANSsec}
Let us now formulate the algorithms. Assume that $L=2^R$: $I=\{L,\ldots,2L-1\}$, $I_s=\{L_s,\ldots,2L_s-1\}$ and that $q_s:=L_s/L\approx p_s$ approximates the probability distribution of the symbols. There are $|I|=2^R$ positions for spreading symbols with $|\{x\in I:\overline{s}(x)=s\}|=L_s$ appearances of symbol $s$. For convenient table handling, we use $X:=x-L\in \{0,\ldots,2^R-1\}$ and store the symbol spread as $symbol[X]\equiv\overline{s}(X+L)$ size $L$ table.

Method~\ref{gen} generates the $decodingTable$ for efficient decoding step from Method~\ref{dec0}. For efficient memory handling while encoding step, the encoding table can be stored in one dimensional form $C(s,x)=encodingTable[x + start[s]] \in I$ for $x\in I_s$, where $start[s]=- L_s + \sum_{s'<s}L_{s'} $.            
To encode symbol $s$ from state $x$, we first need to transfer $k[s]-1$ or $k[s]$ bits, where $k[s] = \lceil lg(L/L_s) \rceil$. This choice can be simplified to $nbBits = (x + nb[s]) >> r$ using a prepared table $nb[]$. Finally, the preparation and encoding step are written as Methods~\ref{encprep} and \ref{enc0} respectively.

\begin{figure*}
    \centering
        \includegraphics[width=14cm]{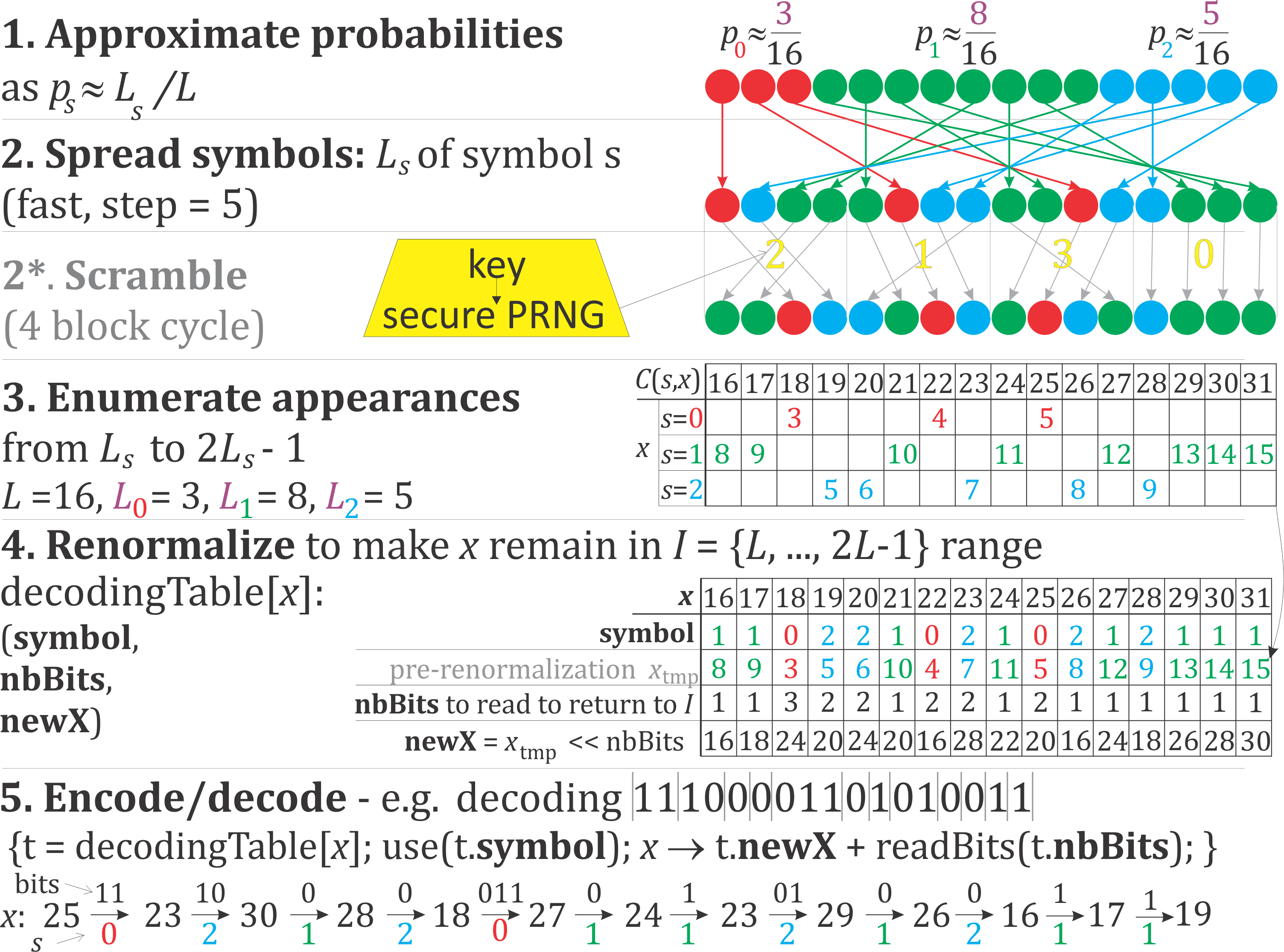}\
\begin{center}
        \caption{Example of generation of tANS tables and applying them for stream decoding for $m=3$ size alphabet and $L=16$ states.}
   \label{ansdiag}
\end{center}
\end{figure*}

\begin{algorithm}[htbp]
\footnotesize{
\caption{Generating tANS $decodingTable$}
\label{gen}
\begin{algorithmic}
\REQUIRE $next[s] = L_s$ \quad\COMMENT{number of next appearance of symbol $s$}
\FOR {$X=0$ to $L-1$}
\STATE $t.symbol=symbol[X]$ \quad\COMMENT{ symbol is from spread function }
\STATE $x=next[t.symbol]++$ \qquad \COMMENT{ $D(X+L)=(symbol,x)$ }
\STATE $t.nbBits = R-\lfloor \lg(x)\rfloor$ \qquad \COMMENT{ number of bits to return to $I$ }
\STATE $t.newX = (x << t.nbBits)-L$\qquad\COMMENT{ properly shift $x$ }
\STATE $decodingTable[X]=t$
\ENDFOR
\end{algorithmic}
}
\end{algorithm}

\begin{algorithm}[htbp]
\footnotesize{
\caption{Preparation for tANS encoding, $L=2^R$, $r=R+1$}
\label{encprep}
\begin{algorithmic}
\REQUIRE $k[s] = R-\lfloor \lg(L_s) \rfloor$  \qquad \COMMENT{$nbBits=k[s]$ or $k[s]-1$}
\REQUIRE $nb[s] = (k[s] << r)-(L_s << k[s])$
\REQUIRE $start[s]= - L_s + \sum_{s'<s}L_{s'}$
\REQUIRE $next[s] = L_s$
\FOR {$x=L$ to $2L-1$}
\STATE $s=symbol[x-L]$
\STATE $encodingTable[start[s] + (next[s]++)] = x$;
\ENDFOR
\end{algorithmic}
}
\end{algorithm}

\begin{spacing}{0.5}
\begin{algorithm}[htbp]
\footnotesize{
\caption{tANS encoding step for symbol $s$ and state $x=X+L$}
\label{enc0}
\begin{algorithmic}
\STATE $nbBits = (x + nb[s]) >> r$    \quad\qquad \COMMENT{$r=R+1,\ 2^r = 2L$}
\STATE useBits$(x, nbBits)$  \qquad\COMMENT {use $nbBits$ of the youngest bits of $x$}
\STATE $x = encodingTable[start[s] + (x >> nbBits)]$
\end{algorithmic}
}
\end{algorithm}
\end{spacing}

\textbf{Symbol Spread Function}:
We need to choose $symbol[X]=\overline{s}(X+L)$ distributing symbols over the $I$ range: $L_s$ appearances of symbol $s$. Finding the optimal way seems is a difficult problem. We present only a fast simple way of spreading symbols in a pseudorandom way in Method~\ref{spread2}, which already offers excellent performance. Several symbol spreads can be found and tested in \cite{toolkit}.

\begin{spacing}{0.2}
\begin{algorithm}[htbp]
\footnotesize{
\caption{Example of fast symbol spread function \cite{fse}}
\label{spread2}
\begin{algorithmic}
\STATE $X=0$; $step=5/8 L +3$ \qquad \COMMENT{ some initial position and step}
\FOR {$s=0$ to $m-1$}
\FOR {$i=1$ to $L_s$}
\STATE $symbol[X]=s$; $X = $mod$(X + step, L)$
\ENDFOR
\ENDFOR
\end{algorithmic}
}
\end{algorithm}
\end{spacing}

\section{Adding encryption}
The construction of tANS code gives us an opportunity to ensure data confidentiality. The concept of encryption in tANS coder is described in this section.

\subsection{Basic concept}
We could use the freedom of choosing the exact coding for encryption purposes. For example while building prefix tree for a size $m$ alphabet, there are $m-1$ internal nodes. Switching their left and right children gives us $2^{m-1}$ options of encoding our message.

As discussed, prefix codes can be viewed as a tANS for $L_s$ being powers of 2 and symbols spread in ranges. Without this restriction, there are much more options of choosing the $\overline{s}$ function:
$${L \choose L_1, \ldots ,L_m} \approx 2^{L\cdot H(L_1/L,\ldots,L_m/L)}$$
where $H(p_1,\ldots,p  _m)=\sum_i p_i \lg(1/p_i)$ is entropy.

Each option defines a different coding. Therefore we need a method of spreading symbols according to the cryptographic key. One way is first to use an independent method, e.g. put successive symbol every $step$ number of positions (cyclically). Then we can perturb the obtained symbol spread using a cryptographically-secure pseudorandom number generator (CSPRNG) seeded with the key, for example by taking blocks and cyclically shifting symbols inside such blocks by a shift from the CSPRNG.

Figure \ref{ansdiag} depicts an example of coding and encryption processes for the following parameters: $L=16$, $m=3$ size alphabet, $step=5$ and size $B=4$ blocks. After step 2, where we spread all symbols (globally), the scrambling process in blocks is performed (locally). This is crucial from the security point of view, since a different locations of symbols results in different forms of encoded messages. The encoded messages depend strongly on the CSPRNG key.

\subsection{Numbers of possibilities}
The key space is a crucial element for protecting the secure cipher against brute-force attacks, therefore we analyze the number of ways of encoding messages.

As default parameters (DP), we consider $L=2048$ states, $m=256$ size alphabet and $B=8$ blocks, which requires 8kB of lookup tables (or 6kB with simple bit compression). As degenerated default parameters (DDP), we consider the worst case scenario: when there is one dominating symbol and the remaining ones have the minimal $L_s=1$ number of appearances.

The number of different symbol spreads for DP is $2^{2048 H}$ and depends on the entropy of the sequence. We can use DDP to find the lower bound: the number of symbol spreads here is $\frac{L!}{(L-m+1)!}\approx 1.65\cdot 10^{837}$ for $L=2048,\ m=256$.

The assumed perturbation using cyclic shifts by values from PRNG reduces these numbers. For DP, this number is $B^{L/B}=8^{256}\approx 1.55\cdot 10^{231}$. Some cyclic shifts of such blocks may accidentally lead to identical symbol alignment. The probability that two $B$ length blocks from the i.i.d. $\{p_i\}$ probability distribution are accidentally equal is approximately $2^{-B H(p_1,\ldots,p_m)}$. Therefore, for practical scenarios (e.g. $m=256$, $H>1$), the reduction of space of possibilities is practically negligible.
For the DDP case, approximately $\left(\frac{L-m+1}{L}\right)^B\approx 0.345$ of blocks have the dominating symbol only. The remaining ones are always changed by the perturbation: the number of possibilities is $\approx B^{(1-0.345) B/L}\approx 2.49\cdot 10^{151}$.

\subsection{Chaotic state behavior}
Having a large number of possible codings is not sufficient; strong dependence on the key is also required. One source is relying on the security of CSPRNG, which ensures that changing a single bit in the key produces a completely independent perturbation of the symbol spread. Additionally, eventual inferring the coding function would give no information about the key (seed).

Another source is chaos of dynamics of the internal state $x$, ensuring that incomplete knowledge leads to a rapid loss of any information about the state of the coder. State $x$ can be viewed as a buffer containing $\lg(x)$ bits of information, and adding a symbol of probability $s$ increases it by $\lg(1/p)$ bits. Due to renormalization, this addition is modulo 1 - accumulated complete bits are send to the stream. Finally the approximate behavior is $\lg(x)\to^\approx \lg(x)+\lg(1/p)\ \mod\ 1$.

\begin{figure}[t!]
    \centering
        \includegraphics[width=8cm]{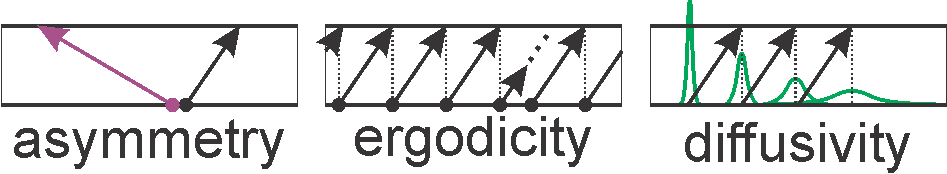}\
\begin{center}
        \caption{Three sources of chaotic behavior of the internal state $x$: $\lg(x)\to^\approx \lg(x)+\lg(1/p)\ \mod\ 1$.}
        \label{chaos}
\end{center}
\end{figure}

\begin{figure}[t!]
    \centering
        \includegraphics[width=8cm]{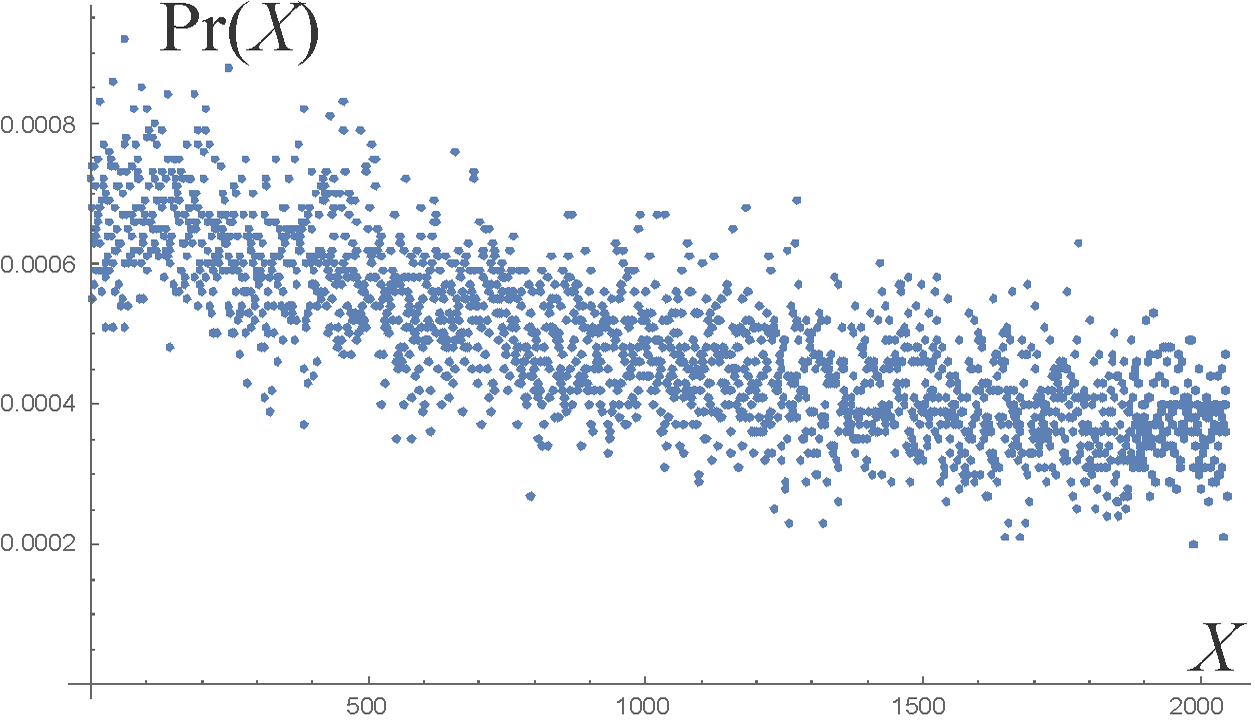}\
\begin{center}
        \caption{Example of probability distribution of $X=x-L$ in $L=2048$ and $m=256$ case and fast symbol spread.}
   \label{dens}
\end{center}
\end{figure}

This cyclic addition formula contains three sources of chosity as depicted in Figure \ref{chaos}:
\begin{itemize}
  \item asymmetry: each position may correspond to a different symbol and so to a different shift,
  \item ergodicity: $\lg(1/p)$ is usually irrational, so even a single symbol tends to cover the range uniformly,
  \item diffusivity: this formula is approximate, so even knowing the symbol sequence, information about the exact position is quickly lost.
\end{itemize}
These properties suggest that we should expect an approximately uniform probability distribution of $lg(x)$, which corresponds to $\Pr(x)\propto 1/x$ distribution of $x$. Better symbol spreads are close to this behavior. For the discussed fast symbol spread, the noise around this $1/x$ curve can be high, as shown in Figure \ref{dens}.

\section{Features and limitations}
The presented concept of lightweight compression with encryption should be verified from the security point of view.
In this section, we discuss results of standard cryptographic tests of tANS encoding with encryption as well as presenting ways to of enhancing the security of this solution. The tests were mainly performed for the DP case: $L=2048$, $m=256$
 for pure tANS layer - imperfections can be easily removed for example by a reduced number of layers of AES.

\subsection{Balancing}
The first question regarding the statistics of the produced bit sequence is the density of ``0" and ``1". Are they equal? In general, for ANS algorithm it is not exactly fulfilled. This is due to the fact that the probability distribution of used states $x$ prefers lower states: approximately $\Pr(x)\propto 1/x$, such as in Figure \ref{dens}. For the DP case, tests show approximately 0.001 difference ($\Pr(0)\approx 0.501$), it has never exceeded $0.002$. For different parameters, an approximate general behavior of this difference is that it is proportional to $m/L$.

For higher correlations, the probability of a length $k$ bit sequence in the produced stream should be $2^{-k}$. Beside the above difference, our tests could not detect further disagreements with this rule.

The variable-length nature of ANS makes the $\Pr(0)\approx 0.501$ issue unlikely to be useful for cryptanalysis (due to the lack of synchronization). Additionally, this small imbalance can be easily removed by adding an inexpensive additional operation, such as XOR with a mask (or set of masks) having equal number of zeros and ones.

\subsection{Avalanche and nonlinearity}
One crucial feature of the secure cipher is the Strict Avalanche Criterion (SAC), which is satisfied if a change of a single bit of the key results, on average, in a change of one half of bits of ciphertext. The tANS approach uses CSPRNG which has a similar property: changing a single bit of the seed leads to a statistically independent random stream, which means an independent tANS coding table. We tested a property which is even stronger than SAC: by encoding the same symbol sequence using the same coding tables, but starting with a different initial state. We were not able to detect statistical dependencies between such two streams.

Additionally, we verified the nonlinearity of the encryption process (defined as the Hamming distance to the nearest affine function). The tests confirmed the nonlinear behavior of encryption process.

\subsection{Diffusion and completeness}
The next important feature is the diffusion of changes during the encryption process. We verified that even when the number of changes in the entry were low, the change of the output bits was high.

The same behavior was observed during the tests of completeness. Completeness is satisfied when a change of a single bit of the plaintext causes a change of around one half of bits of ciphertext. The discussed method processes successive single symbols, so a change of a symbol can influence only bits corresponding to the current and successive positions. We have performed tests with two encoding streams starting with the same state $x$. We first encode a single symbol different for each streams, followed by a sequence of symbols identical for both streams. Encoding a symbol of probability $p$ produces the youngest $\lfloor \lg(1/p)\rfloor$ or $\lceil \lg(1/p)\rceil$ bits of $x$. This means that the first few bits after the change of symbol will be identical -- their number depends on the probability of symbol. Tests of further bits were not able to find statistical dependencies.

Operating on short bit blocks (of varying length) leaves an option for adaptive attacks by exploring ciphertexts differing by single symbols. To protect against this, the initial state can be chosen entirely randomly. We can use such an initial state by analogy to the initial vector in many modes of encryption, e.g. Cipher Block Chaining (CBC).
This way, the same symbol sequences lead to independent bit sequences. As the number of the initial state may be insufficient, this property can be enhanced by adding a few random symbols at the beginning of plaintext.

We can enhance this protection by making sure that we use an independent coding table each time. This can be achieved by using what is referred as 'salt': a random number which affects the seed of CSPRNG and is stored in the header of a file. Additionally, the stream is usually divided into frames of e.g. 10kB size, what is common in data compression applications for updating probability distributions. For encryption purposes, new independent coding tables can be generated for each frame, using the number of frame also as a seed. Finally, we could use triple data: a cryptographic key, the number of the frame and a random number (salt) as the seed of CSPRNG.
\\
\\
Summarizing, the tests of features confirm that presented solution is able to protect confidentiality at a high level of security. We suggest the following principles:
\begin{itemize}
\item using a relatively large number of states and a large alphabet (protecting against brute-force attacks),
\item encrypting the final state, which is required for decoding,
\item using a completely random initial state to protect against adaptive attacks (additionally, appending a few random symbols at the beginning, which are discarded by decoder, would strengthen this protection).
\end{itemize}

During the implementation of proposed solution, it is possible to strengthen the security level by:
\begin{itemize}
\item using three parameters: the cryptographic key, the number of the data frame (e.g. 10kB) and a random number stored in an encrypted file (salt) as a seed for CSPRNG to make all coding tables completely independent,
\item using an inexpensive additional encryption layer, such as XOR with a set of masks (generated using  CSPRNG), a simple substitution-permutation cipher, or AES with a reduced number of rounds.
\end{itemize}

Future work on proposed compression algorithm with encryption process should focus on advanced cryptanalysis and finding the optimal compromise between security and performance.

\section{Conclusion}
This paper proposes a new concept of compression with simultaneous encryption. From the data compression perspective, it provides a nearly optimal compression ratio (such as arithmetic coding) at an even lower cost than Huffman coding (due to having inexpensive linear initialization instead of the $n \log n$ cost of sorting in Huffman coding). Using CSPRNG initialized with a cryptographic key to choose the coding tables means the message encoded this way can be simultaneously encrypted at nearly no additional cost. The variable-length nature of this coding makes eventual cryptanalysis extremely difficult as the attacker does not know how to split the bit sequence into blocks corresponding to successive symbols. These blocks and even their lengths depend on the internal state of the coder, which is hidden from the attacker. The behavior of this state is chaotic, rapidly eliminating any incomplete knowledge of the attacker. Using CSPRNG ensures that even if an attacker would obtain the applied coding table, no information about the cryptographic key is acquired.

Such lightweight compression with encryption is crucial in many situations, for example in battery-powered remote sensors which should transmit the gathered data in a compressed and secure manner. We are entering the age of the Internet of Things, where the use of such types of devices will be widespread. The hundreds of potential applications of this solution include medical implants transmitting diagnostic data, smart RFIDs powered by electromagnetic impulses only, smartphones or smartwatches with improved performance and extended battery life, and many other situations for data storage and transmission.


\bibliographystyle{IEEEtran}
\bibliography{ref}

\begin{thebibliography}{10}
\providecommand{\url}[1]{#1}
\csname url@samestyle\endcsname
\providecommand{\newblock}{\relax}
\providecommand{\bibinfo}[2]{#2}
\providecommand{\BIBentrySTDinterwordspacing}{\spaceskip=0pt\relax}
\providecommand{\BIBentryALTinterwordstretchfactor}{4}
\providecommand{\BIBentryALTinterwordspacing}{\spaceskip=\fontdimen2\font plus
\BIBentryALTinterwordstretchfactor\fontdimen3\font minus
  \fontdimen4\font\relax}
\providecommand{\BIBforeignlanguage}[2]{{%
\expandafter\ifx\csname l@#1\endcsname\relax
\typeout{** WARNING: IEEEtran.bst: No hyphenation pattern has been}%
\typeout{** loaded for the language `#1'. Using the pattern for}%
\typeout{** the default language instead.}%
\else
\language=\csname l@#1\endcsname
\fi
#2}}
\providecommand{\BIBdecl}{\relax}
\BIBdecl

\bibitem{HC}
D.~Huffman, ``A method for the construction of minimum redundancy codes,''
  \emph{Proceedings of the IRE}, vol.~40, no.~9, pp. 1098--1101, September
  1952.

\bibitem{ari}
J.~J. Rissanen, ``Generalized kraft inequality and arithmetic coding,''
  \emph{IBM Journal of research and development}, vol.~20, no.~3, pp. 198--203,
  1976.

\bibitem{ran}
G.~Martin, ``Range encoding: an algorithm for removing redundancy from a
  digitized message,'' \emph{Proceedings of Institution of Electronic and Radio
  Engineers International Conference on Video and Data Recording Conference},
  July 1979, {Southampton, England}.

\bibitem{CABAC}
D.~Marpe, H.~Schwarz, and T.~Wiegand, ``Context-based adaptive binary
  arithmetic coding in the {H.264/AVC} video compression standard,'' \emph{IEEE
  Transactions on Circuits and Systems for Video Technology}, vol.~13, no.~7,
  pp. 620--636, July 2003.

\bibitem{mah}
M.~Mahoney, ``{Data Compression Programs website},''
  {http://mattmahoney.net/dc}.

\bibitem{ans}
J.~Duda, ``Asymmetric numerical systems,'' \emph{arXiv:0902.0271}.

\bibitem{last}
------, ``Asymmetric numeral systems: entropy coding combining speed of huffman
  coding with compression rate of arithmetic coding,'' \emph{arXiv:1311.2540}.

\bibitem{pcs2015}
J.~Duda, K.~Tahboub, N.~J. Gadgil, and E.~J. Delp, ``The use of asymmetric
  numeral systems as an accurate replacement for huffman coding,'' \emph{31st
  Picture Coding Symposium}, 2015.

\bibitem{zhuff}
Y.~Collet, ``Zhuff compressor,''
  \url{http://fastcompression.blogspot.com/p/zhuff.html}.

\bibitem{lzturbo}
H.~Buzidi, ``lzturbo compressor,''
  \url{https://sites.google.com/site/powturbo/}.

\bibitem{LZA}
N.~Francesco, ``Lza compressor,'' \url{http://heartofcomp.altervista.org/}.

\bibitem{LZFSE}
``Apple lzfse compressor,'' \url{https://github.com/lzfse/lzfse}.

\bibitem{ZSTD}
``Facebook zstandard compressor,'' \url{https://github.com/facebook/zstd}.

\bibitem{FPGA}
S.~M. Najmabadi, Z.~Wang, Y.~Baroud, and S.~Simon, ``High throughput hardware
  architectures for asymmetric numeral systems entropy coding,'' in \emph{2015
  9th International Symposium on Image and Signal Processing and Analysis
  (ISPA)}.\hskip 1em plus 0.5em minus 0.4em\relax IEEE, 2015, pp. 256--259.

\bibitem{light1}
T.~Eisenbarth, S.~Kumar, C.~Paar, A.~Poschmann, and L.~Uhsadel, ``A survey of
  lightweight-cryptography implementations,'' \emph{IEEE Design \& Test of
  Computers}, vol.~24, no.~6, pp. 522--533, 2007.

\bibitem{light2}
A.~Y. Poschmann, ``Lightweight cryptography: cryptographic engineering for a
  pervasive world,'' in \emph{Ph. D. Thesis}.\hskip 1em plus 0.5em minus
  0.4em\relax Citeseer, 2009.

\bibitem{light3}
P.~H. Cole and D.~C. Ranasinghe, ``Networked rfid systems and lightweight
  cryptography,'' \emph{London, UK: Springer. doi}, vol.~10, pp. 978--3, 2008.

\bibitem{eLZ1}
D.~Xie and C.-C. Kuo, ``Secure lempel-ziv compression with embedded
  encryption,'' in \emph{Electronic Imaging 2005}.\hskip 1em plus 0.5em minus
  0.4em\relax International Society for Optics and Photonics, 2005, pp.
  318--327.

\bibitem{eLZ2}
J.~Kelley and R.~Tamassia, ``Secure compression: Theory \& practice,''
  Cryptology ePrint Archive, Report 2014/113, 2014.

\bibitem{eBWT}
M.~O. K{\"u}lekci, ``On scrambling the burrows--wheeler transform to provide
  privacy in lossless compression,'' \emph{Computers \& Security}, vol.~31,
  no.~1, pp. 26--32, 2012.

\bibitem{earith1}
I.~H. Witten and J.~G. Cleary, ``On the privacy afforded by adaptive text
  compression,'' \emph{Computers \& Security}, vol.~7, no.~4, pp. 397--408,
  1988.

\bibitem{earith2}
H.~Kim, J.~Wen, and J.~D. Villasenor, ``Secure arithmetic coding,''
  \emph{Signal Processing, IEEE Transactions on}, vol.~55, no.~5, pp.
  2263--2272, 2007.

\bibitem{HCcrypt}
K.-K. Tseng, J.~M. Jiang, J.-S. Pan, L.~L. Tang, C.-Y. Hsu, and C.-C. Chen,
  ``Enhanced huffman coding with encryption for wireless data broadcasting
  system,'' in \emph{Computer, Consumer and Control (IS3C), 2012 International
  Symposium on}.\hskip 1em plus 0.5em minus 0.4em\relax IEEE, 2012, pp.
  622--625.

\bibitem{rivest}
D.~W. Gillman, M.~Mohtashemi, and R.~L. Rivest, ``On breaking a huffman code,''
  \emph{IEEE Transactions on Information theory}, vol.~42, no.~3, pp. 972--976,
  1996.

\bibitem{chaos1}
M.~Baptista, ``Cryptography with chaos,'' \emph{Physics Letters A}, vol. 240,
  no.~1, pp. 50--54, 1998.

\bibitem{chaos2}
G.~Jakimoski, L.~Kocarev \emph{et~al.}, ``Chaos and cryptography: block
  encryption ciphers based on chaotic maps,'' \emph{IEEE Transactions on
  Circuits and Systems I: Fundamental Theory and Applications}, vol.~48, no.~2,
  pp. 163--169, 2001.

\bibitem{ryg}
{F. Giesen}, {https://github.com/rygorous/ryg\_rans}.

\bibitem{toolkit}
{J. Duda}, {https://github.com/JarekDuda/AsymmetricNumeralSystemsToolkit}.

\bibitem{fse}
{Y. Collet}, {https://github.com/Cyan4973/FiniteStateEntropy}.

\end{thebibliography}
\end{document}